# Reduction of Self-heating effect in LDMOS devices

T.K.Maiti[*] and C. K. Maiti[**]

Department of Electronics and Electrical Communication Engineering, Indian Institute of

Technology, Kharagpur-721302, India

Tel: +91-3222-281475, email: tkm.iitkgp@gmail.com

**Abstract**

Isotopic purification of group IV elements leads to substantial increase in thermal conductivity due to reduced scattering of the phonons. Based on this concept, a simulation study is used to demonstrate the reduction of at least 25 $^{o}$C in LDMOS average temperature.

[*]Present address: Hiroshima University, 1–3-1, Kagamiyama, Higashi-Hiroshima City, P.C. 7398530, Japan

[**]Present address: Department of Electronics and Communication Engineering, Institute of Technical Education and Research (ITER), SOA University, Bhubaneswar, 751030, Odisha India

## Introduction

Smart power technologies, integrating high voltage CMOS transistors with standard low voltage CMOS cores, are finding an increasing use in the area of automotive applications, switching power supplies and amplifiers, and devices operating in the industrial environments where the supply voltage busses are in the 12 V to 50 V range. In high voltage applications double-diffused MOS (DMOS) transistors are more attractive than conventional MOS structures in integrated circuits, since the drift region between the drain and the active channel allows high voltage biasing. In contrary to the DMOS, transistor which operates in vertical direction, the LDMOS transistor is lateral orientated and thereby length and width dependent. Currently $0.7 \mu m$ CMOS technology with the addition of 100 V n/p DMOS and 80 V NPN/PNP BJTs is in use. Process technology need to be developed to allow applications up to 100 V. However, in some applications maximum voltages of no more than 40-50 V are required, thus allowing much smaller devices than the existing 100 V devices.

Development of optimal LDMOS devices is a complex process, where several device design parameters need to be taken into account, and simultaneous optimization is needed towards maximum breakdown at minimal $R_{on}$, minimum area, high reliability performance, and ideal output characteristics. The 2005 International Technology Roadmap for Semiconductors (ITRS) states that the use of TCAD will provide as much as a 40% reduction in technology development costs. In today's competitive environment, time to market for new products is of paramount importance. Therefore, a methodology/ strategy is required that reduces the development cycle time. Technology CAD (TCAD) has become one of the key approaches to predict device performance and to optimize process conditions for CMOS development, as it is essential to debug process technologies before manufacturing release. Utilizing TCAD and Design for Manufacturability (DFM) techniques, it is possible to reduce the total cycle time needed in the development of a technology.

Self-heating effect (SHE) in LDMOS results in degradation in performance and reliability. Various techniques are being proposed to reduce the self-heating in MOSFETs. In this paper, we show via simulation that the use of $^{28}$Si isotopes in LDMOS that would reduce the temperature rises in the device due to the reduction in phonon scattering. Using TCAD tools, we have studied the thermal mapping in LDMOS devices fabricated virtually with both the natural Si and isotopically purified silicon ($^{28}$Si) substrates. It is shown that the higher thermal conductivity of $^{28}$Si is useful for the reduction in temperature rise and self-heating. Heat generation and conduction equations are discussed in detail for the study of self-heating-effect in LDMOS.

## Device Description

Sentaurus Sprocess/Sdevice simulators [manual1-2] were used for process development and extracting the simulated electrical parameters. The process flow was developed using Sprocess, and the structure file was used as input into the device simulator Sdevice. The initial process specifications (main process steps) are shown below. The process flow in this work represents a generic LDMOS technology. The major process steps are:

    1. Substrate layer implant and epilayer growth.

    2. Well implants.

3. Well drive-in.

4. LOCOS oxide growth.

5. Post-LOCOS cleanup.

6. Gate oxide growth.

7. Polysilicon gate deposition and reoxidation.

8. Source–drain implants.

9. Source–drain annealing.

10. Body contact implants.

11. Body contact annealing.

12. Metallization.

The simulations were based on models used for the smart power process from which the flow under investigation was derived, therefore, providing some confidence in the usefulness of the models and hence, TCAD tools used. In process simulations a fixed meshing strategy as well as an adaptive meshing strategy was employed. The Sentaurus Device performs device simulations to extract key electrical parameters in order to facilitate customized calibration and optimization. The process model was calibrated by including as much process information in the model as possible.

The process flow was simulated with Sentaurus Process using the equilibrium diffusion model. The two-phase segregation model is activated during LOCOS growth and segregation was neglected. The advantage is that the number of simulation runs to explore the complete parameter space and the resulting device electrical properties can be used to help optimize the process control settings and the analysis of the impact of process parameter sensitivity on device response provides a greater insight into, for example, which process parameter would require monitoring in-line.

Fig. 1 shows the device structure at end of process simulation. Metal regions are gray, oxide regions are brown, and concentrations of dopants are shown in silicon body region and polysilicon gate. The length of the gate on the thin gate-oxide is 2.6 µm, denoted by $L$ and the thick-field-oxide drift region (of length $L_{LOCOS}$) of length 3.5 µm. After each processing step, the simulated intermediate structure is saved. The subsequent processing steps are performed in a new simulation and the previous simulation results are reloaded.

## Heat generation modeling

This work based on the Poisson's equation and electron and hole continuity equations. The Poisson's equation is:

$$\vec{\nabla}.(\varepsilon\vec{\nabla}\psi) = -q(p-n+N_D^+-N_A^-)-\rho_{trap} \qquad (1)$$

Where ε is the electrical permittivity, $q$ is the elementary electronic charge; $n$ and $p$ are the electron and hole densities, $N_D^+$ is the concentration of ionized donors, $N_A^-$ is the concentration of ionized acceptors, and $\rho_{trap}$ is the charge density contributed by traps and fixed charges. The electron and hole continuity equations are,

$$\frac{\partial n}{\partial t} = -\vec{\nabla}.\vec{J}_n + G - R \qquad (2)$$

$$\frac{\partial p}{\partial t} = -\vec{\nabla}.\vec{J}_p + G - R \qquad (3)$$

where $\vec{J}_n$ and $\vec{J}_p$ are particle current densities of electrons and holes; G and R are generation and recombination rates respectively. The heat flow equation under isothermal process is give by,

$$c\left(\frac{\partial T}{\partial t}\right) = \vec{\nabla}.\left(k\vec{\nabla}T\right) + H \quad (4)$$

where, $c$ is the heat capacity, $K$ is the thermal conductivity, and $H$ is the heat generation rate. So Gaur and Navon [GAUR7650] introduced a simplified model for heat generation per unit volume and is,

$$H = \vec{J}.\vec{E} \quad (5)$$

where, $\vec{J}$ is the electric current density and $E$ is the electric field.

A relation between the lateral electric field (E) and the current density in the drift region is given by [KIM911641],

$$E(x) = \left(qN\frac{\mu}{J(x)} - \frac{1}{E_c}\right)^{-1} \quad (6)$$

where, $E_c$ is saturation electric field and is directly involved in the heat generation process. If temperature (T) gradient is exist in the device then eqns. 4 and 5 can be replaced as,

$$\vec{J}_n = n\mu_n\left(\vec{\nabla}\phi_n - \alpha_n\vec{\nabla}T\right) \quad (7)$$

$$\vec{J}_p = -p\mu_p\left(\vec{\nabla}\phi_p - \alpha_p\vec{\nabla}T\right) \quad (8)$$

The coefficients $\alpha_n$, and $\alpha_p$ are the thermoelectric powers associated with the electron and hole system, respectively. Now considering the lattice heat capacity ($C_L$), heat capacity of the electron gas ($C_n$), and heat capacity of hole gas ($C_P$) and also considering the thermal conductivity of semiconductor lattice ($K_L$), conductivity of electron gas ($K_n$), and conductivity of hole gas ($K_p$) we derive following heat conduction equation:

$$C_{tot}\left(\frac{\partial T}{\partial t}\right) = \vec{\nabla}.\left(K_{tot}\vec{\nabla}T\right) + H \tag{9}$$

Where $C_{tot} = C_L + C_n + C_p$ is the total heat capacity, $K_{tot} = K_L + K_n + K_p$ is the total thermal conductivity. Using the eqn. 6 we obtained heat flux equation as,

$$\begin{aligned}H(x) = & q\frac{\left|\vec{J}_n(x)\right|^2}{nN_d\mu_n - \frac{\left|\vec{J}_n(x)\right|}{E_C}} + q\frac{\left|\vec{J}_p(x)\right|^2}{pN_a\mu_p + \frac{\left|\vec{J}_p(x)\right|}{E_C}} \\ & + q(R-G)\left[T\left(\frac{\partial \phi_n}{\partial T}\right)_{n,p} - \phi_n - T\left(\frac{\partial \phi_p}{\partial T}\right)_{n,p} + \phi_p\right] \\ & + qT\left[\left(\frac{\partial \phi_n}{\partial T}\right)_{n,p} - \alpha_n\right]\vec{\nabla}.\vec{J}_n - qT\left[\left(\frac{\partial \phi p}{\partial T}\right)_{n,p} + \alpha_p\right]\vec{\nabla}.\vec{J}_p + qT\left(\vec{J}_n.\vec{\nabla}\alpha_n + \vec{J}_p.\vec{\nabla}\alpha_p\right)\end{aligned} \tag{10}$$

The final H(x) expression is fundamental to solving the heat conduction equation. Eqn. 10 shows that H(x) depends on the external applied voltages. To activate above equation (eqn.10), thermal model is specified in the physics section of the command file and appropriate parameters are defined in the fields of the parameter file.

## Results and Discussion

After the process simulation, Sentaurus Device is used to simulate drain current curves as a function of the drain voltage ($I_d$-$V_{ds}$) for different values of the gate bias. I-V simulation is performed using a two-carrier drift-diffusion transport model. Lattice self-heating effects are included by solving the lattice temperature equation. Appropriate thermal conductivity for silicon defined in the fields of the parameter file. Fig. 2a and 2b shows the simulated temperature profile for different isotopically used silicon substrate of LDMOS devices with uniformly doped drift region. Since, drift region is uniformly doped; heat is generated uniformly inside the drift region. A hotspot is clearly seen near the LOCOS edge. This spot indicates the highest temperature region. The temperature distribution is nearly constant at the bottom of the substrate. Thermal conductivity of a solid is limited by phonon scattering and determined by crystal structure, temperature defects, purity, geometry etc. For high quality samples (where most imperfections have been removed), thermal conductivity due to scattering of phonon by boundary, scattering by impurities and lattice imperfection. Natural silicon is used in Si-CMOS technology. The defect concentration in natural Si, containing 92% $^{28}$Si, 4.7% $^{29}$Si and 3.3% $^{30}$Si, is higher than the each individual isotope of silicon. The impurity concentration in the form of isotope is greater than other defect (e.g., doping) and dominant phonon scattering mechanism that transport heat. It has been reported that the thermal conductivity of natural silicon is 142-148 W/m-K and that for $^{28}$Si 165-227W/m-K (at room temperature) [KIZILYALLI05404]. The device made-up with $^{28}$Si isotope shows nearly 25 °C reduction in temperature within the substrate Fig. 3 shows the variation of temperature for LDMOS simulated with natural Si and $^{28}$Si. The temperature distribution from the drift region to the bottom of the substrate obeys the following fitting equation obtained by us for both type devices.

$$T = T_0 + a * \exp\left(-\frac{x}{L}\right) \tag{11}$$

Where $T$ is the rise of temperature, L is the distance between drift region where hot spot is located and bottom surface of the device, $T_0$ is the maximum temperature at the hot spot, $a$ is the constants. Both type devices follow the same type temperature rise equation. But the temperature rise inside the device is different due to use of silicon substrate with different isotope.

Fig. 4 shows both the simulated and reported experimental drain bias dependence of drain current for LDMO device [3Aarts]. An excellent match between the experiment and simulation is observed which shows the reliability of the predictive simulation using TCAD. Figs. 5 show simulated $I_d$-$V_{ds}$ characteristics for different gate bias. Self heating effect (SHE) is visible over the drain current versus drain voltage ($I_d$-$V_{ds}$) characteristic where it induces the appearance of a negative resistance in the saturation region. Information on temperature increase is important to understanding the failure mechanisms in the device. As the temperature rise increases the drain current due decreases it due to the effective mobility degradation at a high-current region [Canepari]. It shows that the self heating effect is less for $^{28}$Si than natural Si in LDMOS devices. The self heating effect degrades the device performance and cause reliability problems in the gate and source metallization. So, the higher thermal conductivity of $^{28}$Si is useful for the reduction in self-heating and enhance device performances.

**Conclusions**

Thermal modeling for self-heating in LDMOS using TCAD tools has been performed. It is shown via simulation, for the first time, the feasibility of device temperature reduction by using $^{28}$Si substrates. The part of the drift region near the source is found to be much hotter than the other region of the device. This hotter region may lead to performance degradation and reliability problems and should receive attention in device design.

[Canepari] A. Canepari, G. Bertrand, M. Minondo, N. Jourdan, and J.-P. Chante, "DC/HF circuit model for LDMOS including self-heating and quasisaturation," IEEE Research in Microelectronics and Electronics, vol. 2 pp.71-74, 2005.

**Figure captions**

Fig.1. Final device structure at the end of the process simulation.

Fig.2. Temperature distribution with different isotopes: (a) natural Si, and (b) $^{28}$Si.

Fig.3. Comparison of temperature distribution between both type LDMOS devices.

Fig.4. Benchmarking and calibration of simulation results with experimental output characteristics.

Fig.5. Drain current as a function of drain voltage for LDMOS structures simulated for gate biases of 2.4V, 3.4 V and 4.4 V.

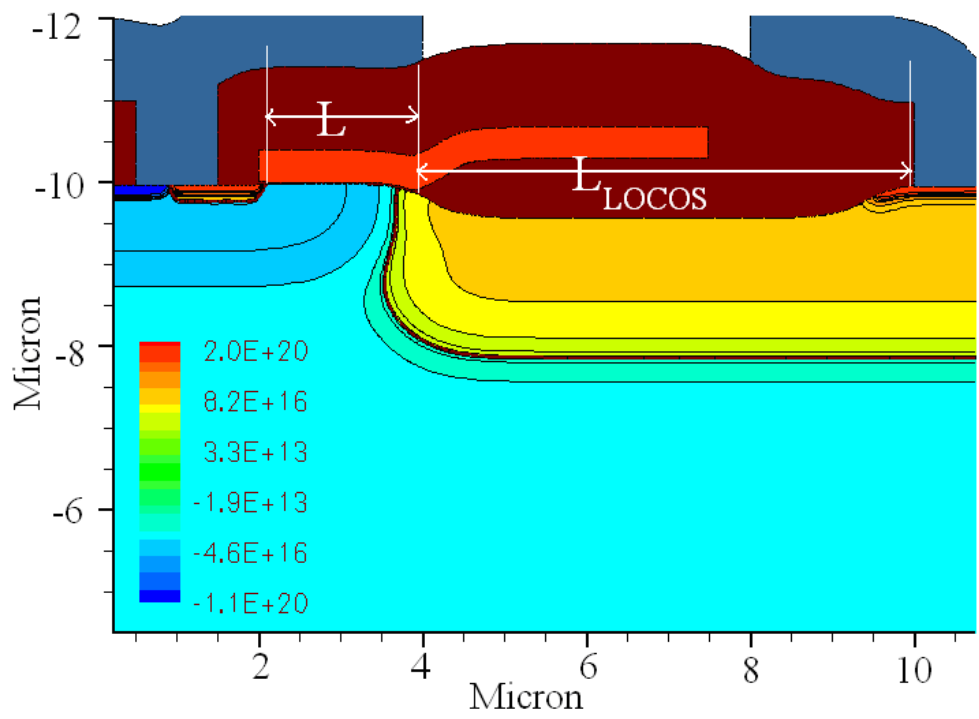

**FIGURE 2**

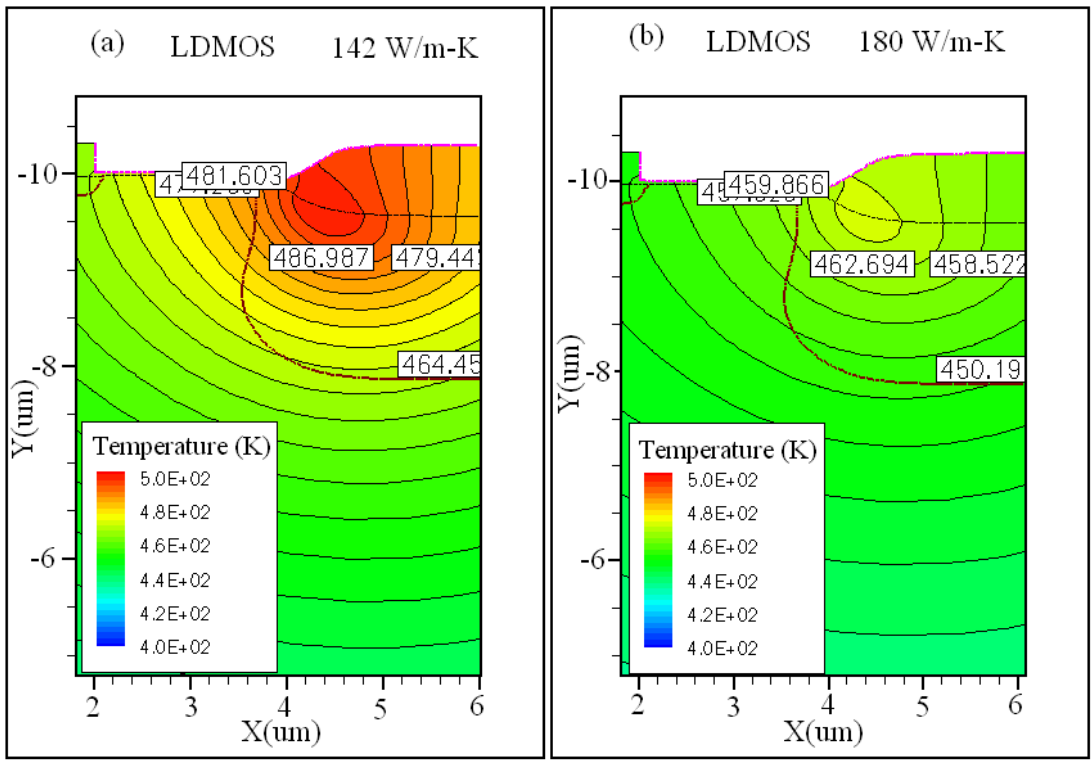

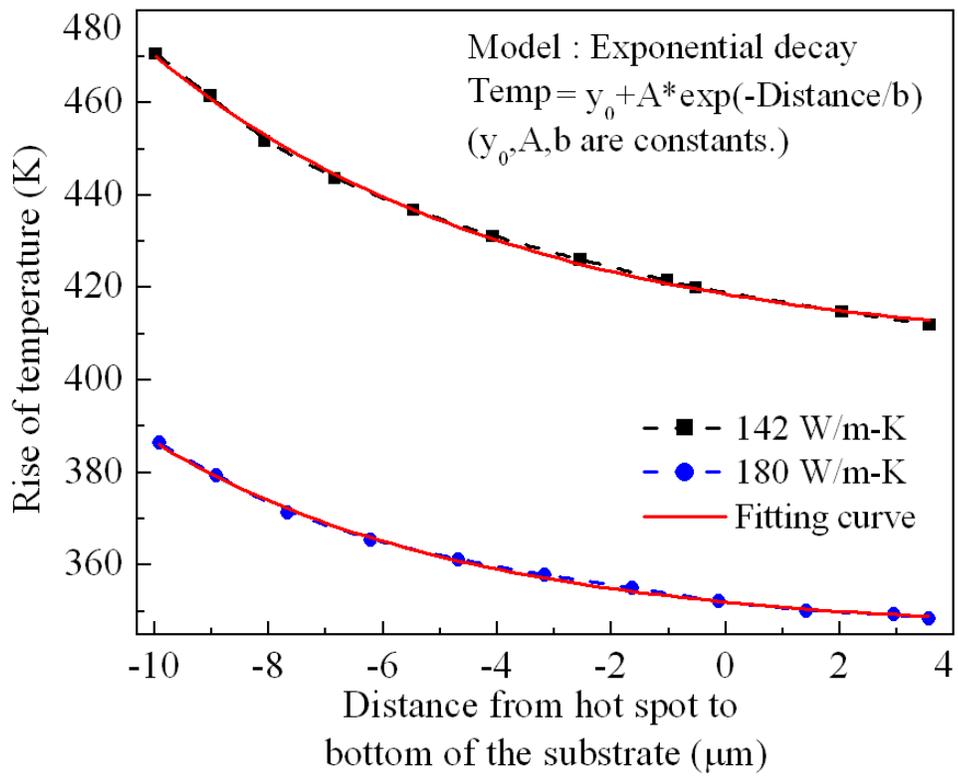

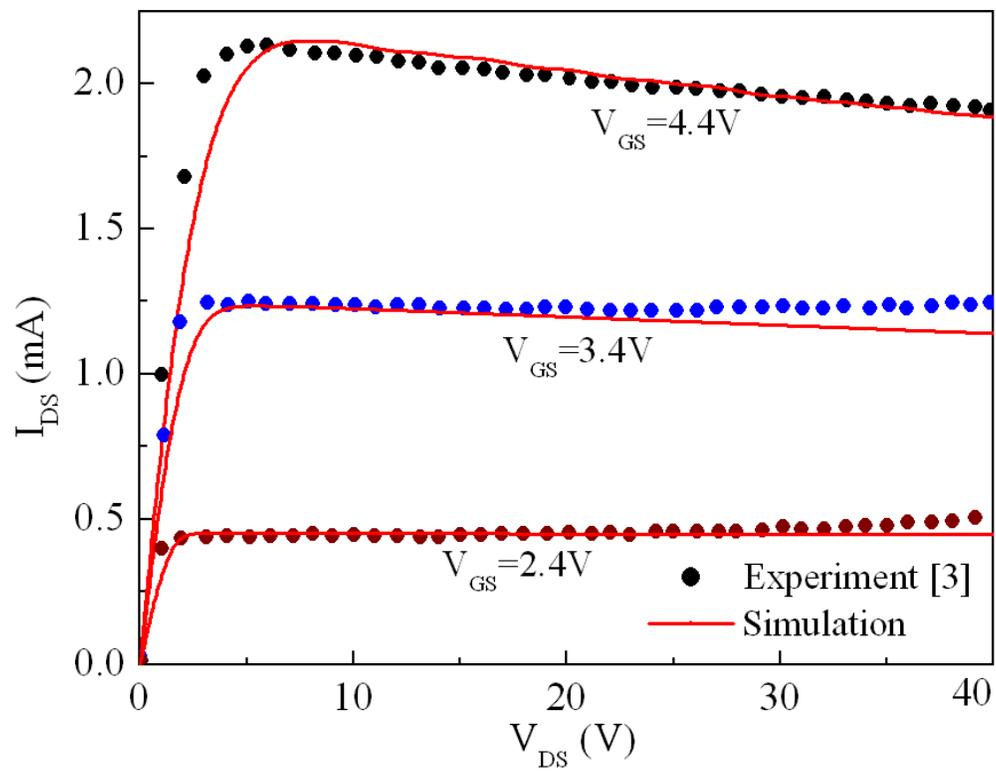

**FIGURE 5**

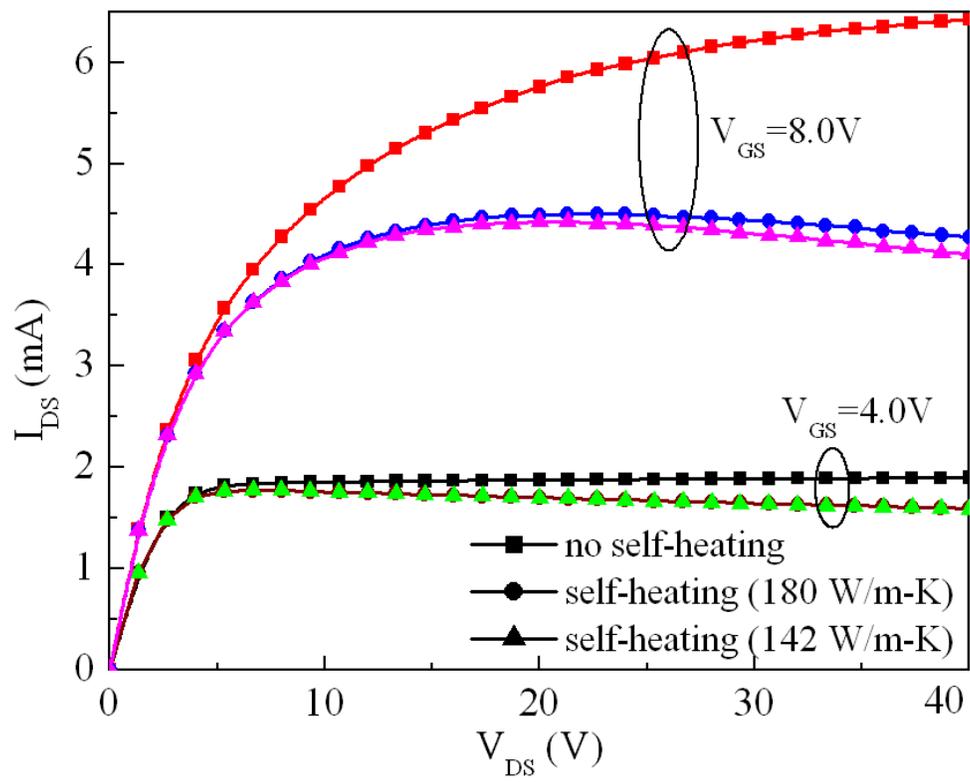